Incremental Propensity Score Effects for Criminology: An Application Assessing the Relationship Between Homelessness, Behavioral Health Problems, and Recidivism

Leah A. Jacobs[1]

Alec McClean[2]

Zach Branson[2]

Edward Kennedy[2]

Alex Fixler[1]

[1] School of Social Work, University of Pittsburgh, Pittsburgh, PA

[2] Statistics and Data Science Department, Dietrich School of Humanities and Social Sciences, Carnegie Mellon University, Pittsburgh, PA

Corresponding Author: Leah A. Jacobs, School of Social Work 2217D Cathedral of Learning, 4200 Fifth Avenue, Pittsburgh, PA 15260, Phone: 412.383.2169.

**Abstract**

**Objectives:** This study examines the relationship between homelessness and recidivism among people on probation with and without behavioral health problems. The study also illustrates a new way to summarize the effect of an exposure on an outcome, the Incremental Propensity Score (IPS) effect, which avoids pitfalls of other approaches commonly used in criminology.

**Methods:** We assessed the impact of homelessness at probation start on rearrest within one year among a cohort of people on probation (n = 2,453). We estimated IPS effects, considering general and crime-specific recidivism if subjects were more or less likely to be unhoused, and assessed effect variation by behavioral health problem status. We used a doubly robust machine learning estimator to flexibly but efficiently estimate effects.

**Results:** A substantial intervention – reducing homelessness by roughly 65% – corresponded to a 9% reduction in the estimated average rate of recidivism ($p < .05$). Milder interventions showed smaller, non-significant effect sizes. Stratifying by behavioral health problem and rearrest type led to similar results without statistical significance.

**Conclusions:** Minding limitations related to observational data and generalizability, this study suggests large reductions in homelessness lead to significant reductions in rearrest rates. Efforts to reduce recidivism should include interventions that make homelessness less likely, but notable differences in recidivism will require these interventions be sizable. Meanwhile, efforts to establish recidivism risk factors should consider alternative effects, like IPS effects, to maximize validity and reduce bias.

**Introduction**

People who are convicted of crimes in the United States (U.S.) are about as likely to recidivate as they are to avoid recidivating (i.e., be rearrested, reconvicted, or have their community supervision revoked; Yukhnenko, Sridhar, et al., 2020). A clear indicator of correctional failure, the commonality of criminal recidivism has motivated an extensive body of research that tests risk and protective factors for reoffending. Scholars and practitioners have placed particular emphasis on establishing dynamic causal risk factors (i.e., those that precede recidivism and, when modified, result in a change in recidivism; Monahan & Skeem, 2014), with the understanding that once established they may be intervened upon to reduce recidivism and ultimately incarceration rates. One such risk factor for recidivism may be homelessness (i.e., the state of lacking a stable place to live).[i] Homelessness is common among criminal legal system-involved people and especially among people with behavioral health problems (i.e., mental or substance use disorder diagnoses), who are also grossly overrepresented in corrections (Couloute, 2018; Jacobs & Gottlieb, 2020; Mulvey & Schubert, 2017).

Despite substantial motivation to establish risk factors for recidivism, like homelessness and behavioral health problems, existing research often relies on statistical models that fail to reflect real world realities (i.e., ecological validity) and that have strong and untestable assumptions that can bias estimates (Vansteelandt et al., 2012). To avoid these problems, we offer a new estimand, the Incremental Propensity Score (IPS), and an efficient doubly robust estimator for it. The Incremental Propensity Score (IPS) effect describes the average outcome if everyone's odds of exposure were multiplied by some factor (see Kennedy, 2019). Put another way, IPS effects characterize what would happen if people were more or less likely to be exposed to some factor (e.g., homelessness) or intervention (e.g., a policy that creates affordable housing) of interest. While analyses often estimate unrealistic effects, exposing everyone versus no one to an intervention (e.g., if all people were homeless versus if none were homeless; or if all received affordable housing versus none received affordable housing), IPS effects measure the effect of more realistic policy interventions, which may decrease the likelihood of homelessness (e.g., via affordable housing development), without eliminating it entirely. Additionally, the assumptions required by the parametric or semi-parametric statistical models commonly used to estimate effects on criminological outcomes can lead to biased estimation. To avoid this bias, we use an efficient nonparametric doubly robust estimator based on the efficient influence function of the IPS effect. This approach allows us to leverage nonparametric statistical methods while achieving fast convergence rates, thereby achieving both flexible and efficient estimation.

In this paper, we aim to advance criminology substantively and methodologically by (1) estimating IPS effects to replicate a prior analysis of the relationship between homelessness and rearrest among a cohort of people on probation and (2) estimating IPS effects in new analyses that test whether the effect of homelessness differs for a potentially particularly vulnerable group– people with behavioral health problems. Prior to our empirical illustration and presentation of findings, we review existing research on housing and recidivism, and further describe IPS effects and our estimation approach.

**Homelessness as a Recidivism Risk Factor**

There are three reasons to consider homelessness a contributor to high recidivism rates. First, affordable housing is in short supply in the U.S. and accessing a stable residence is a common struggle for those involved in criminal legal systems. In the past decade housing costs have risen exponentially, while public investment in affordable housing has stagnated (Rice, 2016; U.S. Census Bureau & U.S. Department of Housing and Urban Development, 1963). As a result, there is insufficient infrastructure to support between 39 and 82 percent of low-income renters, depending on state (Aurand et al., 2022). This shortage interacts with employment obstacles, poverty, and overt discrimination to create significant barriers to stable housing for people involved in criminal legal systems (Couloute, 2018). The rate of homelessness among previously incarcerated individuals is nearly tenfold that of the general public (Couloute, 2018).

Second, theoretically, being unhoused may affect recidivism in a variety of ways. When people are unhoused, they may find shelter or sleep in public and private places, violating municipal codes or trespassing laws (Beckett & Herbert, 2009). People who are unsheltered, or have temporary shelter in places they must vacate during the day, are forced to spend time on the streets and other public places, ultimately rendering them more susceptible to surveillance (Robinson, 2019). The day-to-day realities of housing instability also challenge a person's ability to follow terms of their community supervision (e.g., reporting to probation or parole officers, or keeping track of important documents). Being unhoused may also symbolize a disruption in social connections and control that protect against criminal activity (Clark, 2016).

Third, empirically, studies tend to support a relationship between housing status and recidivism, though existing research is limited and effects are not universal. To date, we know of no true experimental studies on housing status and recidivism, though several quasi-experimental studies exist. One such study by Lutze et al. (2014) examined a multidimensional housing assistance program for people leaving prison in Washington State (n = 416). The authors compared the intervention group to a propensity score matched comparison group in terms of the proportion of recidivism events in each group. They found the intervention group had fewer convictions (21.6% intervention vs. 35.6% Comparison, $p = .002$) and readmissions (37% intervention 56.3% vs. Comparison, $p < .001$), but noted no significant effect on revocations.

Lutze et al.'s study provides evidence for the effect of reentry programs that include housing as a component, speaking to the difference in recidivism rates between those who do and do not participate. However, such analyses do not speak to the effect of housing or homelessness alone, as the intervention was multidimensional.

In order to assess the unique effect of housing circumstances on recidivism, researchers have used observational data on housing status (i.e., where exposure to housing is observed but not manipulated) among people in reentry or under community supervision. For example, in the study above, Lutze et al. (2014) conducted supplementary analyses using administrative data on housing circumstances for their entire sample (i.e., the intervention group and matched comparison group) after the intervention period. Results from multivariate Cox proportional hazard models indicated that homelessness was associated with two to three times greater risk of reconviction, revocation, and readmission, after adjusting for demographics and risk level. Clark's (2016) assessment of first post-prison addresses using multilevel logistic regression models indicated that being released to an emergency shelter was associated with a more than 30% higher odds of rearrest compared to people released to a private residence ($\exp(b) = 1.34$, $p < .05$), but no difference in revocations. In contrast, Harding et al. (2016) used multinomial logit models to assess the relationship between housing circumstances and rearrest among people on parole in Michigan (n = 11,064), finding that, compared to periods living with parents, periods of homelessness were not significantly associated with greater odds of rearrest.

Our own prior research on housing circumstances among people on probation suggested homelessness at probation start significantly increased risk of rearrest (Jacobs & Gottlieb, 2020). Based on Cox proportional hazards models, we found periods of homelessness during probation predicted increased hazards of being rearrested for a property crime (HR = 1.58, $p < .001$), minor crime (HR = 1.54, p < .05), and revocation (HR = 1.78, p < .001), but not violent or drug crimes. We also found homelessness interacts with other risk factors to affect recidivism, with particularly detrimental effects for people who are otherwise low in criminal risk– i.e., but for experiencing homelessness some people on probation would be unlikely to recidivate.

This brief review suggests, though not unilaterally, that homelessness increases recidivism. It also illuminates several gaps and inconsistencies. Few studies assess differences in recidivism between people with and without housing. Instead, studies assess differences in recidivism between people who do and do not participate in multi-dimensional interventions that *include* housing (e.g., Lutze et al., 2014) or between unhoused people and people living in other specific housing circumstances (e.g., living with parents; Harding et al., 2016). Though these studies provide valuable information, they leave policy makers unclear as to the potential benefit of simply creating greater access to housing. Also of interest to policy makers are social problems that affect the most people. Yet, the largest group of system-involved people, those on probation, have received relatively little attention in research on housing and recidivism (Jacobs & Gottlieb, 2020). Because of the rapidity with which people on probation enter and leave detention, results from samples of people released from prison are not necessarily generalizable. Further, existing research does not universally support the link between homelessness and recidivism. Some studies find differing and contradictory estimates of recidivism by re-offense type and sample characteristics (e.g., Clark et al., 2016; Jacobs & Gottlieb, 2020) or suggest no relationship between homelessness and recidivism (Harding et al., 2016). Ultimately, differing operationalizations of homelessness and recidivism, over or underrepresentation of particularly vulnerable subgroups, and statistical modeling decisions may account for inconsistent results. Central to the current study, we address the two latter issues below.

**Homelessness and Recidivism among People with Behavioral Health Problems**

People with behavioral health problems may represent one group at particular risk of recidivism due to homelessness. They are overrepresented in justice systems, and recidivate at similar or greater rates than their relatively well counterparts (Baillargeon et al., 2009; Mulvey & Schubert, 2017; Yukhnenko, Blackwood, et al., 2020). Although people with behavioral health problems generally share risk factors for recidivism (e.g., criminal history, negative social relationships, and anti-social attitudes and personality patterns) with system-involved people *without* such problems (Andrews et al., 2006; Skeem et al., 2011), people with behavioral health problems also face unique factors (e.g., increased surveillance when symptomatic, stigma) likely to increase rearrest and they disproportionately experience homelessness (Balyakina et al., 2014; Herbert et al., 2015; Jacobs et al., 2021; Jacobs & Gottlieb, 2020).

Not only are people with behavioral health problems more likely to experience homelessness, they may be uniquely at risk for recidivism *due to* homelessness. Symptoms may interact with homelessness to garner negative attention from other citizens and law enforcement, in turn increasing the risk of rearrest for people with behavioral health problems. In addition, behavioral health service participation may be a requirement of community supervision for people with behavioral health problems and participation in such services may be challenged by the day-to-day realities of homelessness. If homelessness hinders the ability to attend services as scheduled, then people mandated to such services would be uniquely vulnerable to revocation for violations. Alternatively, behavioral health services could help people deal with the day-to-day realities of homelessness, providing social support and control, and ultimately reducing the effect of homelessness. Despite the logical, though differing, connections between homelessness and rearrest unique to this population and interest in decreasing their overrepresentation in justice systems, we know of no study that has tested whether homelessness differentially affects recidivism for people with behavioral health problems.

**Incremental Propensity Score Effects: An Alternative Approach to Understanding the Relationship between Homelessness and Recidivism**

Previous statistical analyses of the relationship between homelessness and recidivism have two considerable limitations. First, they estimate Average Treatment Effects (ATEs) and Average Treatment Effects on the Treated (ATTs) (Listwan et al., 2018; Nyamathi et al., 2015), which compare relatively extreme interventions – i.e., if everyone versus no one were treated, (for the ATE), and, amongst only those who were treated, everyone versus no one were treated (for the ATT). In this context, this corresponds to estimating the counterfactual average rate of recidivism if everyone versus no one were homeless at the start of probation. Arguably, there are no real-world policies that could approximate these interventions. This is a common problem across the sciences, and there is a new and growing literature proposing solutions (Carneiro et al., 2011; Haneuse & Rotnitzky, 2013).

The Incremental Propensity Score (IPS) effect, first developed by Kennedy (2019) and recently applied in other disciplines (e.g., public health; Naimi et al., 2021), is one approach to summarizing effects that overcomes the limitations described above.[ii] Instead of targeting the "always treated" vs "never treated" contrast, IPS effects describe the average outcome if everyone's odds of treatment were multiplied by some factor. As a result, IPS effects consider what would happen if people on probation were more or less likely to be homeless at the start of probation, both of which could be a consequence of a feasible policy intervention (e.g., increasing or decreasing the number of affordable housing units in a city). This approach provides more information about the effects of various interventions, framing the ATE as a special case, since the "always treated" and "never treated" interventions are two extremes of IPS effects. Furthermore, unlike standard approaches, IPS effects can be efficiently estimated even when some subjects' propensity scores (i.e., probability of experiencing an intervention) are close to zero or one, which often occurs in practice (Westreich & Cole, 2010).

The second limitation of previous analyses is that they assume the relationships between access to housing, recidivism and other predictor variables can be described by simple parametric or semiparametric models. For example, prior analyses have relied on logistic regression models or proportional hazards models (e.g., Harding et al., 2016; Jacobs & Gottlieb, 2020). If these modeling assumptions are incorrect (e.g., the relationship between predictors and the log-odds of recidivating is not linear), then corresponding results can be biased and misleading. Furthermore, standard variable selection procedures for these parametric models, which select predictors based on their association with the outcome, are known to bias intervention effects, where one must consider confounders that are associated with the intervention as well as the outcome (Vansteelandt et al., 2012). Instead, we estimate IPS effects with a "doubly robust" estimator, which uses flexible nonparametric methods for the propensity score and outcome models, and thus relaxes the aforementioned parametric assumptions.

**Current Study**

The current study builds on existing research on the relationship between housing and recidivism, aiming to illustrate an innovative analytic approach, assess the robustness of previously published results to alternative analytic method, and extend analyses to advance substantive knowledge. To do so, we estimate IPS effects with nonparametric double robust estimators using data from a prior study on homelessness and recidivism that relied on Cox proportional hazards models to target the ATE. Specifically, we estimate IPS effects to answer the question: Do changes in the odds of being homeless predict rearrest among people on probation? We hypothesize that, as the odds of being homeless increases, the chance of rearrest will also increase, and we assess potential differences in this effect by offense type. Next, we extend our prior research to explore the relationship between homelessness, behavioral health problems (i.e., psychiatric diagnoses), and rearrest, using IPS effects to assess the potential for homelessness to have a differential effect on people with and without behavioral health problems. Based on our findings, we discuss methodological and substantive implications.

## Methods

### Sample

The sample for this study included 2,453 people on probation from San Francisco, California. We included all people sentenced to standard probation terms who started probation between January 2011 and July 2013.[iii] Because we wanted to control for potential confounders of the relationship between homelessness and recidivism, we excluded 967 people for whom risk assessment data were not available. This yielded an inclusion rate of 72% (2,453 / 3,510).[iv] Most people in the sample are male (85%) and Black (50%), with a mean age of 36 (see Table 1).

**Table 1** Demographic, behavioral health, and recidivism characteristics of the study sample

|  | Housed at Probation Start (n = 1,805) | Homeless at Probation Start (n = 648) | Overall Sample (n = 2,453) |
|---|---|---|---|
| **Age** | | | |
| Mean (SD) | 35.3 (12.1) | 38.7 (11.6) | 36.2 (12.1) |
| **Sex** | | | |
| Female | 293 (16.2%) | 80 (12.3%) | 373 (15.2%) |
| Male | 1,512 (83.8%) | 568 (87.7%) | 2,080 (84.8%) |
| **Race/ethnicity** | | | |
| Black | 903 (50.0%) | 309 (47.7%) | 1,212 (49.4%) |
| Latino | 268 (14.8%) | 77 (11.9%) | 345 (14.1%) |
| Other | 226 (12.5%) | 56 (8.6%) | 282 (11.5%) |
| White | 408 (22.6%) | 206 (31.8%) | 614 (25.0%) |
| **Recidivism Risk** | | | |
| Mean (SD) | -0.132 (1.03) | 0.366 (0.815) | 0.000 (1.00) |
| **Financial Insecurity** | | | |
| Mean (SD) | -0.157 (0.994) | 0.438 (0.879) | 0.000 (1.00) |
| **Diagnosis** | | | |
| Co-occurring | 45 (2.5%) | 55 (8.5%) | 100 (4.1%) |
| No Diagnosis | 1,534 (85.0%) | 453 (69.9%) | 1,987 (81.0%) |
| SMI Only | 102 (5.7%) | 56 (8.6%) | 158 (6.4%) |
| SUD Only | 124 (6.9%) | 84 (13.0%) | 208 (8.5%) |
| **Rearrest Status and Reason** | | | |
| Did not recidivate | 862 (47.8%) | 155 (23.9%) | 1,017 (41.5%) |
| Crime against person | 171 (9.5%) | 56 (8.6%) | 227 (9.3%) |
| Drug crime | 202 (11.2%) | 107 (16.5%) | 309 (12.6%) |
| Minor crime | 107 (5.9%) | 61 (9.4%) | 168 (6.8%) |
| Property crime | 193 (10.7%) | 120 (18.5%) | 313 (12.8%) |
| Violation | 193 (10.7%) | 94 (14.5%) | 287 (11.7%) |
| Unknown | 77 (4.3%) | 55 (8.5%) | 132 (5.4%) |

Notes. Summary statistics represent counts, unless otherwise specified. SD = standard deviation; SMI = Serious mental illness; SUD = Substance use disorder

**Data Sources**

We drew data from probation case records, county court records, behavioral health service records, and comprehensive risk assessments (Correctional Offender Management Profiling for Alternative Sanctions; COMPAS). Probation records contained data on probation start and stop dates, stop reasons, homeless spells, and demographics. County court records contained demographic and rearrest data. Behavioral health service records contained mental and substance use disorder diagnoses. The COMPAS contained data on each person's criminal risk and living situation. The COMPAS was completed during pre-sentence investigations, and includes data collected via probationer self-report and criminal history records (see Measures for further detail). Where data sources provided overlapping demographic or outcome data (i.e., probation stops and rearrests), we verified accuracy by cross-checking records.

**Measures**

The primary outcome in this study is recidivism as indicated by rearrest for a new offense or a violation of probation. We define recidivism as rearrest for either a new crime or a violation within one year of probation start, as have other studies measuring criminal justice failure (Ostermann et al., 2015). We adopted this inclusive definition of recidivism because residential instability and status could compromise probationers' ability to avoid criminal behavior and to uphold community supervision terms. We also constructed offense-specific outcomes to test for variation in the effect of homelessness by offense type. Specifically, we coded offense-specific outcomes as person related crimes, drug crimes, property crimes, minor crimes, or revocations. In doing so, we matched 91% of charge codes to the penal code. We excluded people with unmatchable codes from offense-specific analyses.

Homelessness, our primary predictor of interest, can be measured in a variety of ways. In our prior research, we found that homelessness at probation start is a particularly salient predictor of rearrest. Thus, in this study, we measure homelessness as having no regular place to live at baseline, as indicated via self-report in COMPAS assessments. We verified these reports with probation officer-documented address records. We also explored homelessness over the course of probation, but ultimately did not consider it for this analysis. The fidelity of homelessness records over the course of probation is lower than homelessness at probation start, and the corresponding counterfactual interventions – increasing or decreasing homelessness mid-way through probation – are less feasible than the counterfactual intervention of increasing or decreasing homelessness at the start of probation.

To test for variation in the effect of homelessness between people with and without

INCREMENTAL PROPENSITY SCORE EFFECTS                                                13behavioral health problems, we constructed an indicator based on documented psychiatric disorder diagnoses prior to probation start (diagnoses = 1). Diagnoses included serious mental illnesses (SMI; psychotic, bipolar, and major depressive disorders) and substance use disorders (SUD; excluding nicotine dependence), as documented by clinicians in accordance with the *Diagnostic and Statistical Manual DSM-IV* (American Psychiatric Association, 2000). We examine behavioral health problems, including serious mental and substance use disorders, because both increase risk and they often co-occur (Jacobs et al., 2021). However, because variability could exist within the broad category of psychiatric diagnosis, we also created a four-category factor variable that included "No diagnosis", "Serious mental illness only", "Substance use disorder only", and "Co-occurring serious mental illness and substance use disorder."

To measure the effect of homelessness net of other risk factors for recidivism, we include several variables known to predict recidivism. We include COMPAS general recidivism risk scores. The COMPAS contains age and scales on criminal associates, criminal involvement, criminal personality, criminal thinking, violence, family criminality, vocation/education, finances, non-compliance, leisure/recreation, residential instability, social environment, isolation, socialization, and substance abuse. Several studies have established the predictive validity of the COMPAS, including in this study's sample (see Brennan, Dieterich, & Ehret, 2009, Desmarais, et al., 2016; Farabee et al., 2010; Jacobs & Gottlieb, 2020). Demographics also often predict recidivism (Gendreau et al., 1996; Monahan et al., 2017). Thus, we adjust for age, gender, and ethnoracial group. Age is measured as years since birth at probation start. Ethnoracial group is a categorical variable that includes White (reference), Black, Latino, and Other ethnicity/race. Gender is coded as a binary variable (male = 0).

Finally, since homelessness could partially represent poverty and/or lack of social

INCREMENTAL PROPENSITY SCORE EFFECTS					14support, we include measures of these variables to control for their effects. Specifically, we constructed a social support score based on four items from the COMPAS social isolation scale and constructed a financial insecurity scale based on six items from the COMPAS financial needs scale. These items were not drawn from scales heavily weighted in the COMPAS criminal risk score, and when combined in their respective measures demonstrated good internal consistency.

**Analyses**

Below, we describe the IPS effect and explain and justify our estimation approach. First, we define notation. The data for individual $i$ is $(X_i, A_i, Y_i)$, where $Y_i \in \{0,1\}$ denotes recidivism at the end of one year (1 = recidivated), $A_i \in \{0, 1\}$ denotes housing status at the start of probation (1 = homeless), and $X_i$ denotes covariates. Then, the probability of homelessness at the start of probation (the "propensity score") for people with covariates $X = x$ can be denoted as $\pi(x) = P(A = 1 \mid X)$ and the proportion recidivating within one year of probation start (the "outcome regression") for people with covariates $X = x$ and housing status $A = a$ can be denoted as $\mu(a, x) = E(Y \mid A = a, X = x)$.

The IPS effect is the proportion of subjects recidivating after one year if the odds of homelessness were multiplied by $\delta$; i.e., the mean counterfactual outcome if subjects' odds of treatment were changed from $\frac{\pi(x)}{1-\pi(x)}$ to $\frac{\delta\pi(x)}{1-\pi(x)}$; this results in a propensity score given by $q\{\delta; \pi(x)\} = \frac{\delta\pi(x)}{\delta\pi(x)+1-\pi(x)}$. We denote the IPS effect as $\psi(\delta)$; mathematically, it is

*Equation 1*

$$\psi(\delta) = E\big(q\{\delta; \pi(X)\}\mu(1, X) + [1 - q\{\delta; \pi(X)\}]\mu(0, X)\big).$$

The IPS effect averages over the distribution of the covariates X, mixing the recidivism rate for those homeless $\mu(1, X)$ and housed $\mu(0, X)$, where the mixing fraction is the probability of being homeless $q\{\delta; \pi(X)\}$ under the incremental intervention.

We constructed an estimator for the IPS effect based on its efficient influence function. The efficient influence function is the functional derivative of the estimand in its first-order von Mises expansion, a crucial object in semiparametric efficiency theory (Bickel et al., 1993, van der Laan & Robins, 2003, Hines et al. 2022, Tsiatis, 2006). The efficient influence function of the IPS effect is:

*Equation 2*

$$\varphi(\delta; Z_i) = q\{\delta; \pi(X_i)\}\mu(1, X_i) + [1 - q\{\delta; \pi(X_i)\}]\mu(0, X_i) + \frac{q\{\delta; \pi(X_i)\}}{\pi(X_i)}\{Y_i - \mu(1, X_i)\}$$

$$+ \frac{1 - q\{\delta; \pi(X_i)\}}{1 - \pi(X_i)}\{Y_i - \mu(0, X_i)\} + \frac{\delta\{\mu(1, X_i) - \mu(0, X_i)\}\{A_i - \pi(X_i)\}}{\{\delta\pi(X_i) + 1 - \pi(X_i)\}^2}$$

The efficient influence function in Equation 2 follows a standard form whereby it contains one term which comes directly from the identification of $\psi(\delta)$ in Equation 1 (the first summand), plus three weighted residual terms. Crucially, this form enables estimators based on efficient influence functions to satisfy the favorable property of "double robustness" (i.e., "Neyman orthogonality"; Chernozhukov et al., 2018, van der Laan & Robins, 2003). In this context, double robustness means the error of the IPS effect estimator can be expressed as a product of errors of estimators for the propensity score $\pi$ and outcome regression $\mu$. Therefore, the estimator for the IPS effect can attain parametric convergence rates even when it is constructed with nonparametric estimators for $\pi$ and $\mu$. This is crucial in studies such as ours, where we are not confident a parametric estimator will capture the relationship between the outcome or exposure and covariates, and we wish to instead use flexible nonparametric estimators.

We implemented our estimators using the *ipsi* function in the R package *npcausal* (Kennedy, 2021). To guard against over-fitting nonparametric estimators for $\hat{\pi}$ and $\hat{\mu}$, the estimator uses cross-fitting and trains the estimators on separate data from which they are evaluated. The estimator randomly splits the data in two halves (we focus on two for simplicity, but more than two is also common). On the first half, it estimates models for $\hat{\pi}(X)$ and $\hat{\mu}(A, X)$. Specifically, we use ensemble estimators via the SuperLearner package in R (Polley et al., 2021). On the second half, it calculates

*Equation 3*

$$\hat{\varphi}(\delta; Z_i) = q\{\delta; \hat{\pi}(X_i)\}\hat{\mu}(1, X_i) + [1 - q\{\delta; \hat{\pi}(X_i)\}]\hat{\mu}(0, X_i) + \frac{q\{\delta; \hat{\pi}(X_i)\}}{\hat{\pi}(X_i)} \cdot \{Y_i - \hat{\mu}(1, X_i)\}$$

$$+ \frac{1 - q\{\delta; \hat{\pi}(X_i)\}}{1 - \hat{\pi}(X_i)} \cdot \{Y_i - \hat{\mu}(0, X_i)\} + \frac{\delta\{\hat{\mu}(1, X_i) - \hat{\mu}(0, X_i)\}\{A_i - \hat{\pi}(X_i)\}}{\{\delta\hat{\pi}(X_i) + 1 - \hat{\pi}(X_i)\}^2}$$

which is the estimated efficient influence function value for each individual $i$. Next, the estimator switches the two halves and repeats the process to calculate $\hat{\varphi}(\delta; Z_i)$ for all the individuals in the first half. Finally, the estimate of the IPS effect is the sample average of $\hat{\varphi}(\delta; Z_i)$ over all the data and the variance of this estimate is the sample variance of $\hat{\varphi}(\delta; Z_i)$ over all the data; i.e.,

*Equation 4*

$$\hat{\psi}(\delta) = \frac{1}{n}\sum_{i=1}^{n} \hat{\varphi}(\delta; Z_i), \hat{\sigma}(\delta)^2 = \frac{1}{n-1}\sum_{i=1}^{n}\{\hat{\varphi}(\delta; Z_i) - \hat{\psi}(\delta)\}^2.$$

If the estimators $\hat{\pi}$ and $\hat{\mu}$ satisfy certain rate conditions, an asymptotically valid 95% Wald-style confidence interval for the IPS effect is given by $\hat{\psi}(\delta) \pm 1.96\sqrt{\frac{\hat{\sigma}(\delta)^2}{n}}$.

We repeated the estimator for a grid of $\delta$ values and calculated uniform confidence bands across all $\delta$ values using the multiplier bootstrap. The multiplier bootstrap is a computationally efficient alternative to the nonparametric bootstrap (Belloni et al. 2018, Giné & Zinn 1984; see Kennedy, 2019, Section 4.4 for details). We repeated the analysis with the full sample five times, defining arrest differently each time – specifically, we redefined $Y_i$ as a binary variable indicating whether an individual was rearrested for person-related crimes, drug crimes, property crimes, minor crimes, or revocations over the five iterations. Finally, with any rearrest defined as the outcome, we stratified the analysis by psychiatric diagnosis.

**Estimating the homelessness rate**

To provide context for our results in the next section, we also estimated the homelessness rate under different interventions (i.e., at different values of δ). Mathematically, the estimand is

$$\theta(\delta) = E[q\{\delta; \pi(X)\}],$$

which is the homelessness rate if everyone's odds of homelessness were multiplied by $\delta$. We use the same efficient influence function-based approach described above to construct a doubly robust estimator $\hat{\theta}(\delta)$. Specifically, the efficient influence function is $\frac{\delta\{A-\pi(X)\}}{\{\delta\pi(X)+1-\pi(X)\}^2}$ and the doubly robust estimator is

*Equation 5*

$$\hat{\theta}(\delta)_{doublyrobust} = \frac{1}{n}\sum_{i=1}^{n} q\{\delta; \hat{\pi}(X_i)\} + \frac{\delta\{A - \hat{\pi}(X_i)\}}{\{\delta\hat{\pi}(X_i) + 1 - \hat{\pi}(X_i)\}^2}$$

We also implemented the estimator using sample splitting and the SuperLearner to estimate $\hat{\pi}$.

## Results

We present results below through four figures, corresponding to study aims. Figure 1 speaks to the relationship between housing status at probation start and any rearrest within one year of the start of probation in the entire sample. Figure 2 focuses on the entire sample but explores the relationship between housing status and offense-specific rearrest. Figure 3 stratifies results from the first analysis, illustrating effects for subsamples with and without psychiatric diagnoses, and Figure 4 stratifies further, illustrating effect for subsamples with substance use, mental illness, and co-occurring diagnoses. We find the SuperLearner greatly improves balance in the data (see Appendix A), suggesting that the estimators capture confounding.

**Do Changes in the Probability of Being Homeless Predict Recidivism among People on Probation?**

More than a quarter of people (26.42%; n = 648) in the study sample (n = 2,453) lacked access to housing at probation start (see Table 1). Figure 1 shows that reducing the probability of homelessness reduces the estimated rate of recidivism; specifically, it shows that if the likelihood of homelessness were decreased, the estimated average rate of recidivism after one year decreases, and vice versa if the likelihood of homelessness were increased. The x-axis shows $\delta$, the multiplier on the odds of homelessness at the start of probation. Homelessness, with $\delta = 1$ corresponds to the intervention where subjects' probability of homelessness remains unchanged. The y-axis shows the estimated counterfactual average rate of recidivism one year after the start of probation with the shifted probability of homelessness. The grey band is a uniform 95% confidence interval over all $\delta$.

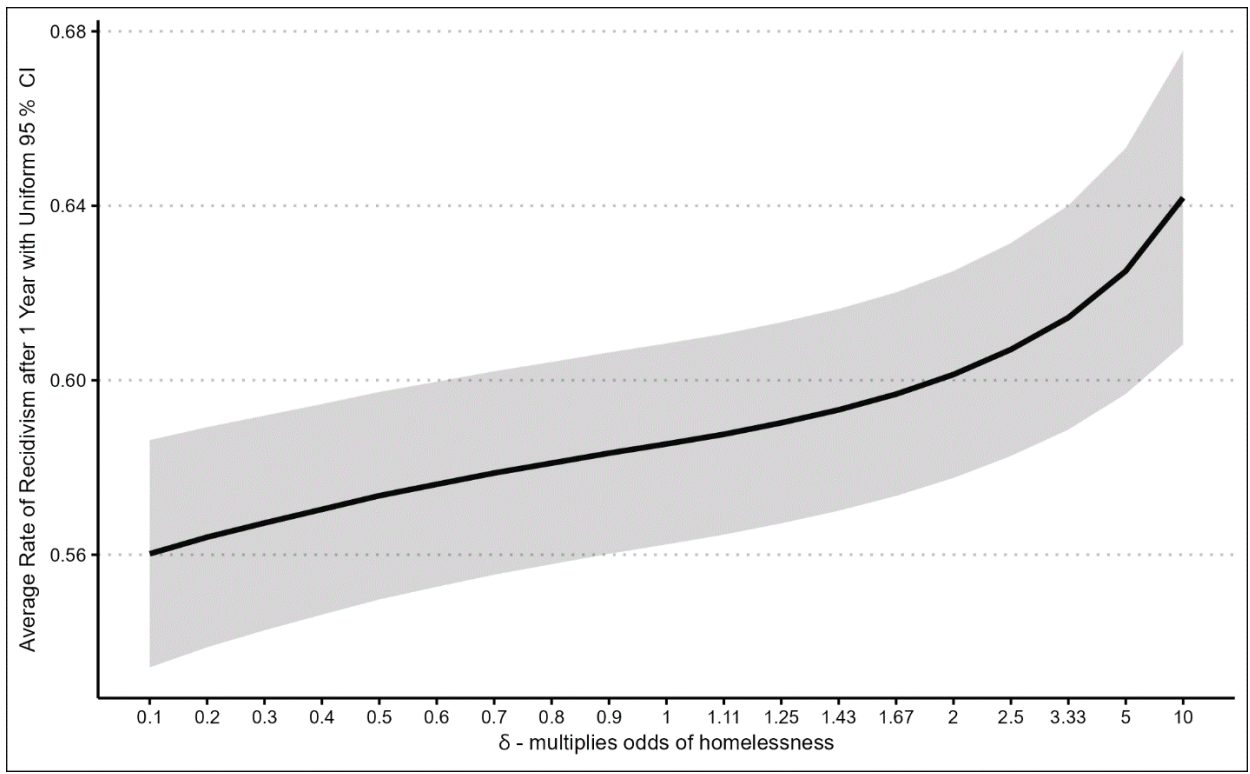

**Fig. 1** Incremental Propensity Score effect curves for average rate of 1-year recidivism (black curve, y-axis) when odds of homelessness at the start of probation are multiplied by different factors ($\partial$, x-axs), with a 95% uniform confidence interval (grey band)

Under the status quo (at $\delta = 1$), we estimate that the average rate of recidivism is roughly 58%. If the odds of homelessness were divided by ten (at $\delta = 0.1$), we estimate the rate of recidivism would decrease to roughly 56%, and if the odds of homelessness were multiplied by ten (at $\delta = 0.1$), the rate of recidivism would increase to 65%. The implied (harmful) treatment effect between these two interventions is nearly a 10% increase in the rate of recidivism. The effect (comparing $\delta = 0.1$ and $\delta = 10$) is significant at the $\alpha = 0.05$ level – we tested the null hypothesis that the rate of recidivism is equal at $\delta = 0.1$ and $\delta = 10$ by checking whether confidence intervals at $\delta = 0.1$ and $\delta = 10$ overlap and rejected the null hypothesis since they do not overlap. However, for milder interventions, such as comparing $\delta = 0.5$ and $\delta = 2$, the effect is smaller and not statistically significant at the $\alpha = 0.05$ level.

We also estimated the homelessness rate to quantify the size of different interventions. Under the status quo (at $\delta = 1$), the homelessness rate is the observed rate, 26%. If the odds of homelessness were halved (at $\delta = 0.5$), the homelessness rate would decrease to roughly 16%, and if the odds of homelessness were doubled (at $\delta = 2$), the homelessness rate would increase to roughly 40%. If instead the odds of homelessness were divided by 10 ($\delta = 0.1$), the homelessness rate would decrease to roughly 5%, while if the odds of homelessness were multiplied by 10 ($\delta = 10$), the homelessness rate would increase to roughly 70%. Therefore, the intervention that compares $\delta = 2$ and $\delta = 0.5$ reduces the homelessness rate by roughly 24%, while the intervention that compares $\delta = 10$ and $\delta = 0.1$ reduces the homelessness rate by roughly 65%. These results and Figure 1 suggest that the effect of homelessness on rearrest is small for what may be considered feasible shifts in the homelessness rate.

Figure 2 shows IPS effects for different re-offense types. There is an inverse trend for Crime Against Person, whereby the likelihood of rearrest increases with decreased probability of homelessness. For the other four arrest types, decreasing probability of homelessness leads to lower rates of recidivism and increasing homelessness leads to higher rates of recidivism. However, the primary takeaway is that all the confidence intervals overlap and thus none of the offense specific tests yield statistically significant results.

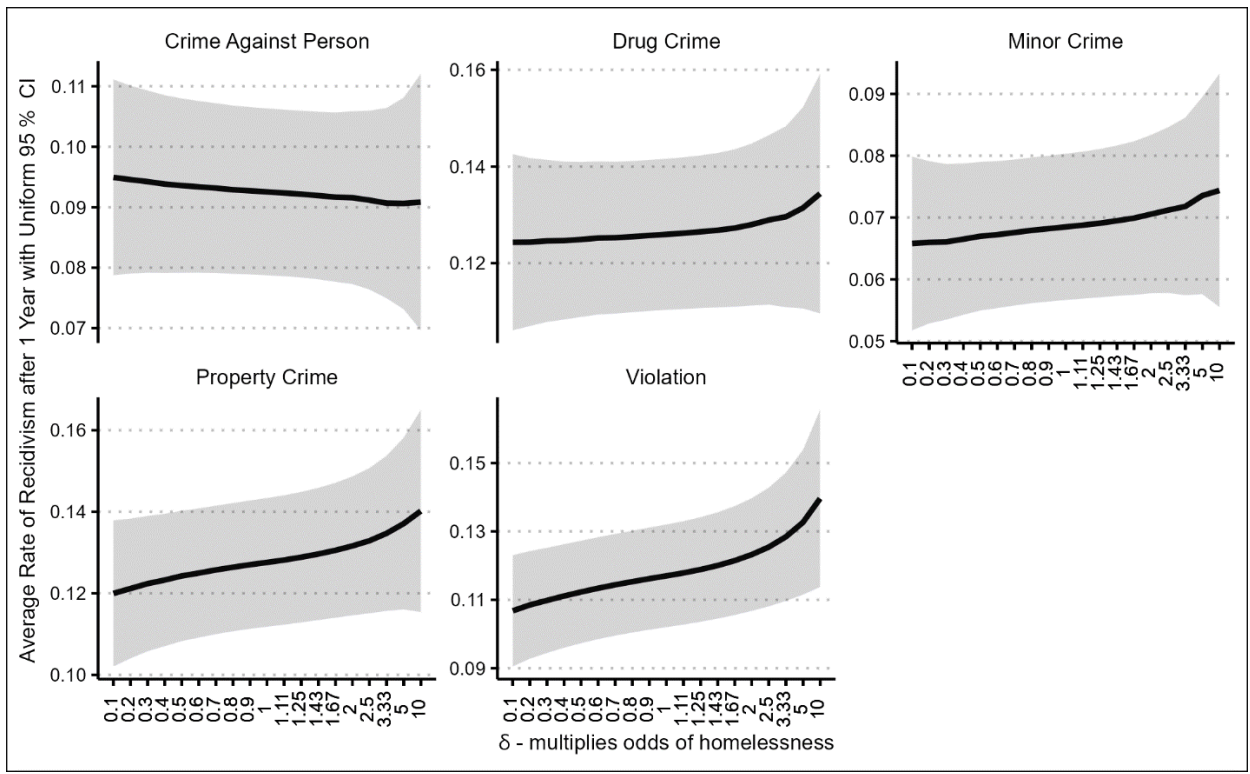

**Fig. 2** Incremental Propensity Score effect curves for average rate of 1-year recidivism (black curve, y-axis) when odds of homelessness at the start of probation are multiplied by different factors (∂, x-axs), with a 95% uniform confidence interval (grey band); separate curves shown for different re-offense types

**Does Homelessness Have a Differential Effect on People with and without Behavioral Health Problems?**

People with a diagnosed behavioral health problem in the sample experienced relatively high risk of homelessness and recidivism. The rate of homelessness among people with behavioral health problems was nearly double that of their relatively well counterparts (41.8% vs. 22.8%) and the rate of recidivism among people with behavioral health problems was about 72% compared to about 55% for those without behavioral health problems. Like Figures 1 and 2, Figure 3 suggests that increased homelessness predicts increased recidivism for people with and without behavioral health problems, though that interpretation should be made with caution. For those without any diagnosis, we see a similar figure to Figure 1, where decreasing homelessness leads to lower rates of recidivism and increasing homelessness leads to higher rates of recidivism. For the group with any diagnoses, we see a generally similar trend, but the confidence bands are very wide.

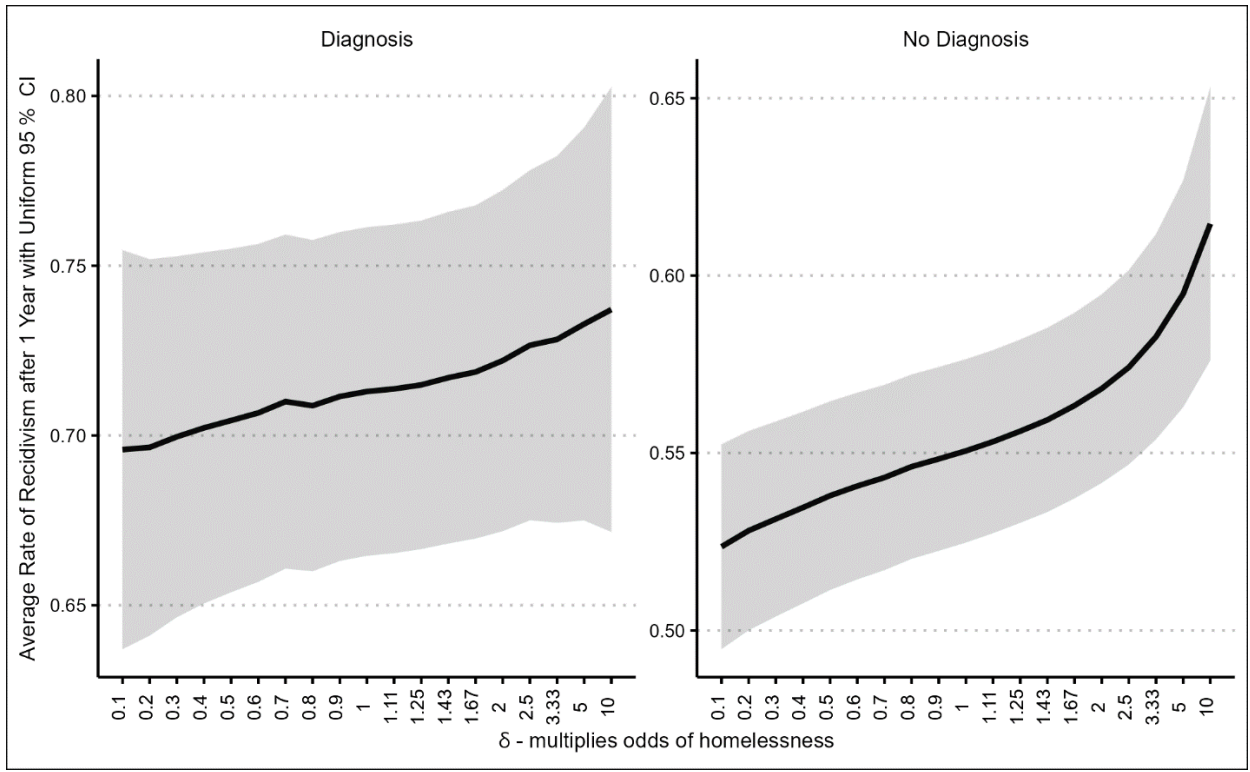

**Fig. 3** Incremental Propensity Score effect curves for average rate of 1-year recidivism (black curve, y-axis) when odds of homelessness at the start of probation are multiplied by different factors ($\partial$, x-axs), with a 95% uniform confidence interval (grey band); separate effect curves are shown for people with and without behavioral health diagnosis

One possible explanation for the lack of a statistically significant effect for those with diagnosed behavioral health problems is that there is too much variation in risk for recidivism within this subgroup (i.e., with some subgroup members affected more than others). Another possible explanation is that, like our exploratory analysis of specific offense types, the study is underpowered to detect effects for those with diagnoses. To assess the first possibility, we explored variation by diagnostic group (see Co-occurring, SMI Only, and SUD Only panels in Figure 4). There were notable differences between these groups in terms of effect sizes, with the clearest apparent effect on those with SUD only. Though all trended toward decreased risk of recidivism when homelessness decreases and vice versa, none of the specific diagnostic categories yielded statistically significant results. This is likely because the study is under-powered to detect effects within these sub-groups; there are 100, 158, and 208 people in the Co-occurring, SMI Only, and SUD Only groups respectively, but 1,987 in the group without diagnoses, a ten-fold increase. In sum, the lack of a statistically significant relationship between homelessness and recidivism among those with any behavioral health problem is likely a reflection of both within group variation and insufficient sample size.

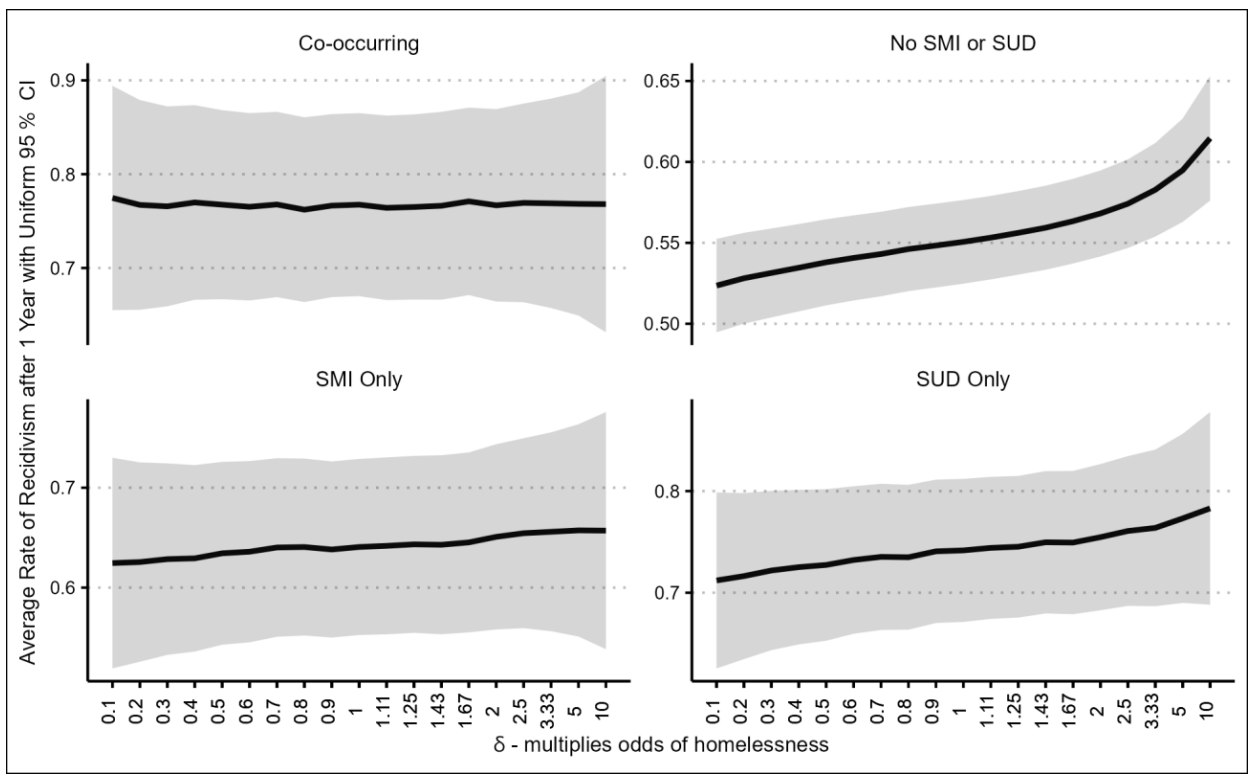

**Fig. 4** Incremental Propensity Score effect curves for average rate of 1-year recidivism (black curve, y-axis) when odds of homelessness at the start of probation are multiplied by different factors ($\partial$, x-axs), with a 95% uniform confidence interval (grey band); separate effect curves are shown for each behavioral health problem (SMI = Serious mental illness, SUD = Substance use disorder, Co-occurring = Both SMI and SUD)

## Discussion

Recidivism is a major challenge to community corrections, with the majority of criminal legal system-involved people failing to avoid further system involvement. Overcoming this challenge requires research that can accurately identify risk factors for recidivism and estimate the potential effects of interventions that intervene upon those risk factors to promote successful community reentry. This paper contributes to such a body of research by introducing a novel counterfactual estimand, the Incremental Propensity Score (IPS) effect, and an efficient nonparametric doubly robust estimator for it. The paper also advances knowledge on the relationship between housing status and criminal recidivism by testing this relationship for a sub-population at significant risk for recidivism—people with behavioral health problems. In the remainder of the paper, we unpack our findings, assess our approach of estimating IPS effects, and draw conclusions for intervention and research.

Similar to prior research, we found homelessness and recidivism were common (Couloute, 2018; Harding et al., 2016). In our sample, more than one quarter (26%) of people on probation had no regular place to live at the start of their probation term and more than half (58%) were rearrested within a year. The preponderance of homelessness and recidivism was even more dramatic among people with behavioral health problems, 42% of whom had no regular housing at probation start and nearly three quarters (72%) of whom were rearrested within one year.

As for homelessness as a risk factor for recidivism, we assessed the relationship between homelessness at the start of probation and the likelihood of rearrest one year later. To do so, we estimated IPS effects with a nonparametric doubly robust estimator. We found that decreasing homelessness may reduce the rate of rearrest; specifically, if the observed odds of homelessness were divided by ten versus if the odds of homelessness were multiplied by ten, the average rate of recidivism after one year would decrease by nearly 10% ($p < 0.05$). However, we observed that this intervention corresponds with a large change in the homelessness rate (from 70% to 5%), which may be unrealistic, and milder interventions did not result in significant effects. This suggests that, without very large reductions in homelessness, change in the rate of rearrest will likely be minimal. These findings align with those from our prior research (Jacobs & Gottlieb, 2020) in that they too indicate homelessness is associated with increased risk of recidivism. However, by indicating strong and sizable interventions would be necessary to see modest reductions in recidivism, results here also temper interpreting our prior study's results as evidence that housing provision is a panacea for recidivism reduction.

    Finally, we find people with behavioral health problems disproportionally experience homelessness and rearrest, but find no evidence they are unique vulnerability to homelessness as it affects rearrest. Analyzing variation in the effect of homelessness on recidivism by behavioral health problem status and type, we found effects were small and similar in direction across groups. In our prior research, we found the effect of homelessness on rearrest particularly weak for people on probation with relatively high overall risk of recidivism (Jacobs & Gottlieb, 2020). Our findings here align, demonstrating that homelessness has a small effect on recidivism among those with behavioral health problems (who experience relatively high overall risk for recidivism). Ultimately, even if people with behavioral health problems are modestly affected and similarly vulnerable to the effects of homelessness as their relatively well counterparts, high rates of homelessness and recidivism among people with behavioral health problems suggests that directing multidimensional interventions, inclusive of housing, toward this group will likely yield benefits for recidivism reduction.

In our view, the use of IPS effects taken here offers an advantage above standard strategies for understanding the relationship between risk factors and criminological outcomes. In the context of homelessness and recidivism, the standard approach has typically involved estimating ATEs or ATTs with parametric regression or semi-parametric hazard modeling of the effect homelessness or changes in housing status on recidivism (e.g., Harding et al., 2016; Steiner et al., 2015), accounting in some manner for potential confounders. While good reasons exist for harnessing observational methods to understand the relationship between homelessness and recidivism (e.g., it would be unethical to randomly assign someone to homelessness), problems exist with standard approaches. Among these problems is that the ATE and ATT assess extreme interventions. The ATE and ATT consider the effect of giving housing to or withholding housing from all people on probation or all people with housing at the start of probation (or some other population of interest) -- extreme and impractical interventions. In research terms, this approach to estimation is a threat to ecological validity.

In contrast, IPS effects build on a relatively reasonable conceptualization of intervention. In the real world, interventions are rarely deployed in the manner of a controlled trial, where one group of people would be completely housed and another group would be completely denied housing. Instead, people may be offered participation in a housing program, people may intermittently access housing, or policies and economic conditions may affect the availability of housing. In other words, the probability of homelessness may be increased or decreased. By estimating the counterfactual average rate of recidivism if everyone's odds of homelessness were multiplied or divided by some factor, IPS uses a milder and more realistic counterfactual intervention than the "always treated" or "never treated" scenarios considered by the ATE and ATT. Further, estimating IPS effects with a doubly robust estimator, which has strong statistical guarantees under weak assumptions, also alleviates concerns about statistical modeling bias.

The abovementioned study results should be placed in the context of three key limitations. One limitation of our approach, and any drawing on observational data, is that it relies on there being no unobserved potential confounders, which can only be established by experimental design. Since experiments are infeasible for studying this problem, we have attempted to reduce the risk of confounding by adjusting for a set of key covariates, including a comprehensive recidivism risk score and social and economic factors. Future analyses could also conduct sensitivity analyses to see how much violation of the no unmeasured confounding assumption might affect results. A second limitation is that sample sizes were small for particular subgroups of subjects or types of rearrest, thereby limiting the statistical power of our subgroup analyses. A third limitation of our approach pertains to generalizability. The study sample represents a group of people on probation with relatively serious index offenses and relatively high risk of recidivism. This is at least in part due to state policy efforts that have encouraged county responsibility for justice-involved persons who might receive prison and parole terms in other states. Further, San Francisco faces one of the most significant affordable housing shortages in the U.S., presenting a particularly challenging circumstance for legal system-involved residents to access housing. Thus, results are most appropriately generalized to samples with similar risk profiles and geographies with similar housing contexts.

**Conclusion**

Across populations and re-offense types, this study indicates increased homelessness tends to predict increased risk of recidivism. Thus, results from this study support efforts to reduce recidivism that include increasing access to housing, including interventions for people on probation and for people with or without behavioral health problems. However, this study also indicates strong and sizable housing interventions are necessary for meaningful reductions in recidivism. Methodologically, efforts to establish recidivism risk factors should consider the appropriateness of traditional approaches that estimate counterfactual outcomes in scenarios unlike real world interventions. To maximize ecological validity and reduce potential bias, researchers should consider estimating alternative effects like Incremental Propensity Score effects.

---

[i] Authors acknowledge that the term "homelessness" may not be the most accurate descriptor for the phenomenon of interest, as advocates have indicated with their turn toward the term "houselessness.". However, as "homeless" is the federally recognized term and remains a common term, we use "homelessness" in lieu of "houselessness."

[ii] Although IPS effects describe a curve of counterfactual average outcomes rather than a contrast between counterfactual average outcomes under different interventions, we refer to the values along this curve as *effects* to maintain consistency with the prevailing literature on the topic. A comprehensive review of IPS effects is provided in Bonvini et al. (2023).

[iii] Standard people on probation are defined as those who would typically be sentenced to supervision by county probation. Standard probationers excludes those who are supervised by probation under California's Public Safety Realignment law, who have committed more serious felony offenses and would traditionally be supervised under state parole.

[iv] Comparison of those included and excluded indicated that those included were comparable to those excluded in terms of age and gender, but those included were more likely to be Black ($x^2$= 39.54, p < .001) and to recidivate ($x^2$= 265.94, p < .001).

**References**

American Psychiatric Association. (2000). *Diagnostic and Statistical Manual of Mental Disorders, text revision (DSM-IV-TR)*. American Psychiatric Association.

Andrews, D. A., Bonta, J., & Wormith, J. S. (2006). The recent past and near future of risk and/or need assessment. *Crime & Delinquency*, *52*(1), 7–27. https://doi.org/10.1177/0011128705281756

Aurand, A., Emmanuel, D., Clarke, M., Rafi, I., & Yentel, D. (2022). *The gap: A shortage of affordable homes* (p. 32). National Low Income Housing Coalition. https://nlihc.org/gap

Baillargeon, J., Binswanger, I. A., Penn, J. V., Williams, B. A., & Murray, O. J. (2009). Psychiatric disorders and repeat incarcerations: The revolving prison door. *American Journal of Psychiatry*, *166*(1), 103–109. https://doi.org/10.1176/appi.ajp.2008.08030416

Balyakina, E., Mann, C., Ellison, M., Sivernell, R., Fulda, K. G., Sarai, S. K., & Cardarelli, R. (2014). Risk of future offense among probationers with co-occurring substance use and mental health disorders. *Community Mental Health Journal*, *50*(3), 288–295. https://doi.org/10.1007/s10597-013-9624-4

Beckett, K., & Herbert, S. (2009). *Banished: The new social control in urban America*. Oxford University Press.

Belloni, A., Chernozhukov, V., Chetverikov, D., & Wei, Y. (2018). Uniformly valid post-regularization confidence regions for many functional parameters in Z-estimation framewok. *Annals of Statistics, 46*(6B), 3643–3675. https://doi.org/10.1214/17-AOS1671

INCREMENTAL PROPENSITY SCORE EFFECTS                                                    30
Bickel, P. J., Klaassen, C. A., Ritov, Y. A., & Wellner, J. A. (1993). Efficient and adaptive estimation for semiparametric models (Vol. 4). Baltimore: Johns Hopkins University Press.

Bonvini, M., McClean, A., Branson, Z., & Kennedy, E. H. (2023). Incremental causal effects: An introduction and review. In J.R. Zubizarreta, E.A. Stuart, D.S. Small & P.R. Rosenbaum (Eds.), *Handbook of Matching and Weighting Adjustments for Causal Inference* (pp. 349-372). Chapman and Hall/CRC.

Brennan, T., Dieterich, W., & Ehret, B. (2009). Evaluating the predictive validity of the COMPAS risk and needs assessment system. *Criminal Justice and Behavior*, *36*(1), 21–40. https://doi.org/10.1177/0093854808326545

Carneiro, P., Heckman, J. J., & Vytlacil, E. J. (2011). Estimating marginal returns to education. *American Economic Review*, *101*(6), 2754–2781. https://doi.org/10.1257/aer.101.6.2754

Chernozhukov, V., Chetverikov, D., Demirer, M., Duflo, E., Hansen, C., Newey, W., & Robins, J. (2018). Double/debiased machine learning for treatment and structural parameters. *The Econometrics Journal, 21*(1), C1-C68. https://doi.org/101111/ectj.12097

Clark, V. A. (2016). Predicting two types of recidivism among newly released prisoners. *Crime & Delinquency*, *62*(10), 1364–1400. https://doi.org/10.1177/0011128714555760

Couloute, L. (2018). *Nowhere to Go: Homelessness among formerly incarcerated people*. Prison Policy Initiative. https://www.prisonpolicy.org/reports/housing.html

Desmarais, S. L., Johnson, K. L., & Singh, J. P. (2016). Performance of recidivism risk assessment instruments in U.S. correctional settings. *Psychological Services*, *13*(3), 206–222. http://dx.doi.org/10.1037/ser0000075

Farabee, D., Zhang, S., Roberts, R. E. L., & Yang, J. (2010). COMPAS validation study: Final report. Semel Institute for Neuroscience and Human Behavior.

Gendreau, P., Little, T., & Goggin, C. (1996). A meta-analysis of the predictors of adult offender recidivism: What works! *Criminology*, *34*(4), 575–608. https://doi.org/10.1111/j.1745-9125.1996.tb01220.x

Giné, E., & Zinn, J. (1984). Some limit theorems for empirical processes. *The Annals of Probability*, *12*(4), 929-989. https://doi.org/10.1214/aop/1176993138

Haneuse, S., & Rotnitzky, A. (2013). Estimation of the effect of interventions that modify the received treatment. *Statistics in Medicine*, *32*(30), 5260–5277. https://doi.org/10.1002/sim.5907

Harding, D. J., Morenoff, J. D., Dobson, C. C., Lane, E. B., Opatovsky, K., Williams, E.-D. G., & Wyse, J. (2016). Families, prisoner reentry, and reintegration. In L. M. Burton, D. Burton, S. M. McHale, V. King, & J. Van Hook (Eds.), *Boys and men in African American families* (pp. 105–160). Springer International Publishing. https://doi.org/10.1007/978-3-319-43847-4_8

Herbert, C. W., Morenoff, J. D., & Harding, D. J. (2015). Homelessness and housing insecurity among former prisoners. *The Russell Sage Foundation Journal of the Social Sciences: RSF*, *1*(2), 44–79. https://doi.org/10.7758/rsf.2015.1.2.04.

Hines, O., Dukes, O., Diaz-Ordaz, K., & Vansteelandt, S. (2022). Demystifying statistical learning based on efficient influence functions. *The American Statistician, 76*(3), 292-304. https://doi.org/10.1080/00031305.2021.2021984

Jacobs, L. A., Fixler, A., Labrum, T., Givens, A., & Newhill, C. (2021). Risk factors for criminal recidivism among persons with serious psychiatric diagnoses: Disentangling what matters for whom. *Frontiers in Psychiatry*, *12*, 778399.

https://doi.org/10.3389/fpsyt.2021.778399

Jacobs, L. A., & Gottlieb, A. (2020). The effect of housing circumstances on recidivism: Evidence from a sample of people on probation in San Francisco. *Criminal Justice and Behavior*, *47*(9), 1097–1115. https://doi.org/10.1177/0093854820942285

Kennedy, E. H. (2019). Nonparametric causal effects based on incremental propensity score interventions. *Journal of the American Statistical Association*, *114*(526), 645–656. https://doi.org/10.1080/01621459.2017.1422737

Kennedy, E. H. (2021). *npcausal: Nonparametric causal inference methods in R* (0.1.0) [Computer software].

Listwan, S.J., Hartman, J.L., LaCourse, A. (2018). Impact of the MeckFUSE Pilot Project: Recidivism among the chronically homeless. *Justice Evaluation Journal 1*(1), 96-108. https://doi.org/10.1080/24751979.2018.1478236

Lutze, F. E., Rosky, J. W., & Hamilton, Z. K. (2014). Homelessness and reentry: A multisite outcome evaluation of Washington state's reentry housing program for high risk offenders. *Criminal Justice and Behavior*, *41*(4), 471–491. https://doi.org/10.1177/0093854813510164

Monahan, J., & Skeem, J. L. (2014). Risk redux: The resurgence of risk assessment in criminal sanctioning. *Federal Sentencing Reporter*, *26*(3), 158–166. https://doi.org/10.1525/fsr.2014.26.3.158

Monahan, J., Skeem, J., & Lowenkamp, C. (2017). Age, risk assessment, and sanctioning: Overestimating the old, underestimating the young. *Law and Human Behavior*, *41*(2), 191–201. https://doi.org/10.1037/lhb0000233

INCREMENTAL PROPENSITY SCORE EFFECTS                                                                    33Moore, S. C., Patel, A. V., Matthews, C. E., Berrington de Gonzalez, A., Park, Y., Katki, H. A., Linet, M. S., Weiderpass, E., Visvanathan, K., Helzlsouer, K. J., Thun, M., Gapstur, S. M., Hartge, P., & Lee, I.-M. (2012). Leisure time physical activity of moderate to vigorous intensity and mortality: A large pooled cohort analysis. *PLoS Medicine*, *9*(11), e1001335. https://doi.org/10.1371/journal.pmed.1001335

Mulvey, E. P., & Schubert, C. A. (2017). Mentally ill individuals in jails and prisons. *Crime and Justice*, *46*(1), 231–277. https://doi.org/10.1086/688461

Naimi, A. I., Rudolph, J. E., Kennedy, E. H., Cartus, A., Kirkpatrick, S. I., Haas, D. M., ... & Bodnar, L. M. (2021). Incremental Propensity Score Effects for Time-fixed Exposures. *Epidemiology*, *32*(2), 202-208.

Nyamathi, A.M., Zhang, S., Salem, B.E., Farabee, D., Hall, B., Marlow, E., Faucette, M., Bond, D., & Yadav, K. (2016). A randomized clinical trial of tailored interventions for health promotion and recidivism reduction among homeless parolees: outcomes and cost analysis. *Journal of Experimental Criminology, 12*(1), https://doi.org/10.1007/s11292-015-9236-9

Ostermann, M., Salerno, L. M., & Hyatt, J. M. (2015). How different operationalizations of recidivism impact conclusions of effectiveness of parole supervision. *Journal of Research in Crime and Delinquency*, *52*(6), 771–796. https://doi.org/10.1177/0022427815580626

Polley E., LeDell, E., Kennedy, C., Lendle, S., van der Laan, M., (2021). *SuperLearner (Version 2.0-28) (https://cran.r-project.org/web/packages/SuperLearner/index.html)*

Rice, D. (2016). *Chart book: Cuts in federal assistance have exacerbated families' struggles to afford housing*. Center on Budget and Policy Priorities.

https://www.cbpp.org/research/housing/cuts-in-federal-assistance-have-exacerbated-families-struggles-to-afford-housing

Robinson, T. (2019). No right to rest: Police enforcement patterns and quality of life consequences of the criminalization of homelessness. *Urban Affairs Review*, *55*(1), 41–73. https://doi.org/10.1177/1078087417690833

Skeem, J. L., Manchak, S., & Peterson, J. K. (2011). Correctional policy for offenders with mental illness: Creating a new paradigm for recidivism reduction. *Law and Human Behavior*, *35*(2), 110–126. https://doi.org/10.1007/s10979-010-9223-7

Steiner, B., Makarios, M. D., & Travis, L. F. (2015). Examining the effects of residential situtions and residential mobility on offender recidivism. *Crime & Delinquency*, *61*(3), 375–401. https://doi.org/10.1177/0011128711399409

Tsiatis, A. A. (2006). *Semiparametric theory and missing data*. Springer.

U.S. Census Bureau & U.S. Department of Housing and Urban Development. (1963, January 1). *Average sales price of houses sold for the United States*. FRED, Federal Reserve Bank of St. Louis; FRED, Federal Reserve Bank of St. Louis. https://fred.stlouisfed.org/series/ASPUS

van der Laan, M. J., & Robins, J. M. (2003). *Unified methods for censored longitudinal data and causality*. Springer.

Vansteelandt, S., Bekaert, M., & Claeskens, G. (2012). On model selection and model misspecification in causal inference. *Statistical Methods in Medical Research*, *21*(1), 7–30. https://doi.org/10.1177/0962280210387717

Westreich, D. & Cole, S. R. (2010). Invited commentary: positivity in practice. *American Journal of Epidemiology*, *171*(6):674–677. https://doi.org/10.1093/aje/kwp436

Yukhnenko, D., Blackwood, N., & Fazel, S. (2020). Risk factors for recidivism in individuals receiving community sentences: A systematic review and meta-analysis. *CNS Spectrums*, *25*(2), 252–263. https://doi.org/10.1017/S1092852919001056

Yukhnenko, D., Sridhar, S., & Fazel, S. (2020). A systematic review of criminal recidivism rates worldwide: 3-year update. *Wellcome Open Research*, *4*, 28. https://doi.org/10.12688/wellcomeopenres.14970.3

**Appendix A: Balance Table**

|  |  | Unadjusted Standardized Mean Difference | Weighted Standardized Mean Difference |
|---|---|---|---|
| **Age** | Continuous | -0.2915 | -0.0175 |
| **Sex** | Binary | 0.1112 | -0.0580 |
| **Race/ethnicity** |  |  |  |
| Black | Binary | 0.0469 | 0.0175 |
| Latino | Binary | 0.0872 | 0.0072 |
| Other | Binary | 0.1264 | -0.0652 |
| White | Binary | -0.2076 | 0.0199 |
| **Recidivism Risk** | Continuous | -0.5371 | -0.0404 |
| **Financial Insecurity** | Continuous | -0.6336 | -0.0488 |
| **Diagnosis** |  |  |  |
| Co-occurring | Binary | -0.2655 | -0.0291 |
| No Diagnosis | Binary | 0.3668 | 0.0344 |
| SMI Only | Binary | -0.1163 | -0.0133 |
| SUD Only | Binary | -0.2049 | -0.0139 |
| **Effective Sample Size** |  |  |  |
| Homeless |  | 648 | 369.9 |
| Housed |  | 1805 | 1650.2 |

Notes. Standardized mean differences are housed minus homeless. SD = standard deviation; SMI = Serious mental illness; SUD = Substance use disorder

The table shows traditional balance diagnostics. The first column shows the variable in the data while the second column shows the type of the variable (continuous or binary, in this case). The third column shows the unadjusted mean differences between the housed and homeless groups. For binary variables, this is the raw difference in proportions, while for continuous variables it is the standardized mean difference. The fourth column shows the same data but re-weighted according to the propensity score estimates. Specifically, the weights for individual $i$, which we denote by $W_i$, are:

*Equation 6*

$$W_i = \frac{A_i}{\hat{\pi}(X_i)} + \frac{1 - A_i}{1 - \hat{\pi}(X_i)}$$

where $A_i = 1$ if individual $i$ was housed at the started of probation and 0 otherwise, and $\hat{\pi}(X_i)$ are the estimated propensity score values for individual $i$. The final two rows of the table show the true and weighted sizes for the housed and homeless groups.

The table demonstrates that balance is improved by weighting according to the estimated propensity scores, indicating that the SuperLearner estimator for the propensity score is appropriately capturing confounding in the observed data.